\documentclass[11pt]{article}
\usepackage{comment}
\usepackage[hidelinks]{hyperref}
\usepackage{enumitem}
\usepackage{amsmath,amssymb,epsf,cite,graphicx,subfigure}
\usepackage{amsmath,braket,tikzsymbols}
\usepackage[bbgreekl]{mathbbol}
\usepackage{comment}
\usepackage{soul}
\numberwithin{equation}{section}

\definecolor{airforceblue}{rgb}{0.36, 0.54, 0.66}
\newcommand{\beq}{\begin{equation}}
\newcommand{\eeq}{\end{equation}}

\begin{document}
\baselineskip=15.5pt
\pagestyle{plain}
\setcounter{page}{1}
%--------+---------+---------+---------+---------+---------+---------+

\begin{center}
{\LARGE \bf AdS$_3$ wormholes from a modular bootstrap}
\vskip 1cm

\textbf{Jordan Cotler$^{1}$ and Kristan Jensen$^{2}$}

\vspace{0.5cm}

{\it ${}^1$ Society of Fellows, Harvard University, Cambridge, MA 02138 \\}

\vspace{0.3cm}

{\it ${}^2$ Department of Physics \& Astronomy, San Francisco State University, \\ San Francisco, CA 94132 \\}

\vspace{0.1cm}

{ \tt  ${}^a$jcotler@fas.harvard.edu, ${}^b$kristanj@sfsu.edu\\}

\medskip

\end{center}

\vskip1cm

\begin{center}
{\bf Abstract}
\end{center}
\hspace{.3cm}
In recent work we computed the path integral of three-dimensional gravity with negative cosmological constant on spaces which are topologically a torus times an interval. Here we employ a modular bootstrap to show that the amplitude is completely fixed by consistency conditions and a few basic inputs from gravity. This bootstrap is notably for an ensemble of CFTs, rather than for a single instance. We also compare the 3d gravity result with the Narain ensemble. The former is well-approximated at low temperature by a random matrix theory ansatz, and we conjecture that this behavior is generic for an ensemble of CFTs at large central charge with a chaotic spectrum of heavy operators.

\newpage

\tableofcontents

%------------------------------------------
\section{Introduction}
%------------------------------------------

In~\cite{Cotler:2020ugk} we obtained the path integral of three-dimensional gravity with negative cosmological constant on spaces that are topologically a torus times an interval. These spaces are Euclidean wormholes, which smoothly connect two asymptotic regions with torus conformal boundaries. Those tori have independent complex structures $\tau_1$ and $\tau_2$. The result was
\begin{align}
\begin{split}
\label{E:mainResult}
	Z_{\mathbb{T}^2\times I}(\tau_1,\bar{\tau}_1,\tau_2,\bar{\tau}_2) &= \frac{1}{2\pi^2}Z_0(\tau_1,\bar{\tau}_1)Z_0(\tau_2,\bar{\tau}_2) \sum_{\gamma\in PSL(2;\mathbb{Z})} \frac{\text{Im}(\tau_1)\text{Im}(\gamma \tau_2)}{|\tau_1+\gamma \tau_2|^2}\,,
	\\
	Z_0(\tau,\bar{\tau})&=\frac{1}{\sqrt{\text{Im}(\tau)}|\eta(\tau)|^2}\,,
\end{split}
\end{align}
where $\gamma \tau = \frac{a\tau+b}{c\tau+d}$, $ad-bc =1$, and $\eta(\tau)$ is the Dedekind eta function. 

It is difficult to interpret Euclidean wormholes within the standard AdS/CFT framework~\cite{Maldacena:2004rf}. Inspired by the duality~\cite{Saad:2019lba} between Jackiw-Teitelboim gravity and a random matrix ensemble, we~\cite{Cotler:2020ugk} (see also~\cite{Afkhami-Jeddi:2020ezh,Maloney:2020nni,Belin:2020hea,Maxfield:2020ale}) have made the working hypothesis that pure 3d gravity is indeed a consistent theory of quantum gravity, dual to an ensemble of conformal field theories (CFTs). Under that hypothesis, the connected ensemble average of torus partition functions is equated with a sum over Euclidean geometries which connect two asymptotic regions with torus boundary,
\beq
	\includegraphics[width=4.5in]{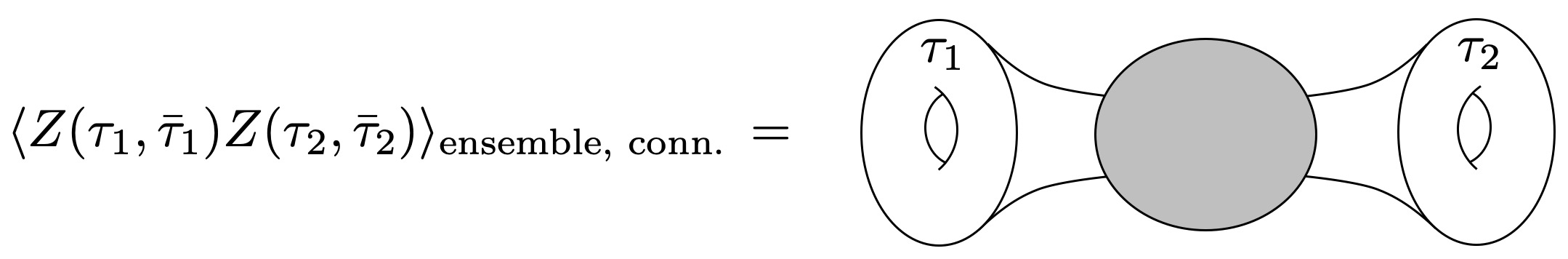}
\eeq
or, in an equation,
\beq
\label{E:ZZ}
	\langle Z(\tau_1,\bar{\tau}_1)Z(\tau_2,\bar{\tau}_2)\rangle_{\text{ensemble, conn.}} = Z_{\mathbb{T}^2\times I}(\tau_1,\bar{\tau}_1,\tau_2,\bar{\tau}_2) + \cdots\,,
\eeq
where the dots refer to other connected geometries with more complicated topology.  Pure 3d gravity does not have a coupling constant which suppresses fluctuations of topology, and so a priori we do not know if the torus times interval contribution dominates. However, at low temperature and fixed large spin $s$~\cite{Ghosh:2019rcj}, pure 3d gravity has an effective Jackiw-Teitelboim description with an effective genus expansion parameter $g_s \sim e^{-2\pi \sqrt{c|s|/6}}$ where $c=\frac{3}{2G}$ is the Brown-Henneaux central charge. So at low temperature and fixed large spin, the torus times interval amplitude gives the dominant contribution to~\eqref{E:ZZ}. Thus, if a dual to pure 3d gravity exists, then the torus partition function is a random variable, and the dual is an ensemble average.

There is a great deal of physics in the wormhole amplitude~\eqref{E:mainResult}. See~\cite{Cotler:2020ugk} for a discussion. In brief, the amplitude describes the two-point fluctuation statistics of BTZ black hole microstates. Near threshold, these match a universal prediction from random matrix theory with Virasoro symmetry.

It was quite tricky to compute $Z_{\mathbb{T}^2\times I}$\,, largely because pure 3d gravity on the torus times interval has no saddle point to expand around. To arrive at~\eqref{E:mainResult} we employed a ``constrain first'' approach, viewing the torus times interval as the annulus times a circle, which admits a Lorentzian continuation to the annulus times time. Pure 3d gravity on the annulus has a phase space of configurations that solve the Gauss' Law constraints, and (up to one subtlety) we performed a path integral quantization of that phase space. The subtlety is that one of the directions of the phase space is a relative time translation between the two boundaries. In Lorentzian signature this time twist has an infinite field range, but in Euclidean signature we redefined it so that it corresponds to a relative Euclidean time translation. The redefined twist has a compact field range.

However, the ``constrain first'' approach is not guaranteed to work (although, as we will see, it does work in our case). That is, quantization does not always commute with constraining. For example, one can~\cite{Elitzur:1989nr} ``constrain first'' to obtain the partition function of Chern-Simons theory on the annulus times a circle, a Wess-Zumino-Witten partition function, which is determined by symmetry and modular invariance. But while the ``constrain first'' approach can give the correct result for the spectrum of a Chern-Simons theory, it can fail to give the correct inner product~\cite{Axelrod:1989xt}.

In the language of path integrals, the basic problem is that when quantizing gauge theory one fixes a gauge and includes the appropriate Faddeev-Popov ghosts. Integrating out the time component of the gauge field enforces the Gauss' Law constraint, which may differ from the classical one by a ghost bilinear. So the residual path integral obtained by quantizing, then constraining, may differ from the one obtained by constraining, then quantizing. Only the former is guaranteed to give the correct answer.

In this paper, we provide a streamlined derivation of~\eqref{E:mainResult} using consistency conditions rather than detailed knowledge of the gravitational path integral. We show that~\eqref{E:mainResult} is completely fixed by Virasoro symmetry and modular invariance after using some basic inputs from 3d gravity. These considerations fix $Z_{\mathbb{T}^2\times I}$ up to an overall normalization, which in turn is determined by the Jackiw-Teitelboim limit of 3d gravity~\cite{Ghosh:2019rcj}. Equivalently, the normalization is fixed by demanding that the two-point spectral correlations are well-described near the spectral edge by random matrix theory.

Another way of saying this is that, after using some facts from gravity, the leading contribution to $\langle Z(\tau_1,\bar{\tau}_1)Z(\tau_2,\bar{\tau}_2)\rangle_{\text{ensemble, conn.}}$ is uniquely fixed by a modular bootstrap.  Notably, this is a modular bootstrap for an ensemble of CFTs, not a single instance of a CFT. Indeed there has been some hope that -- after importing lessons from gravity in AdS space like large central charge, a large twist gap, etc. -- the bootstrap program can identify candidate CFTs dual to theories of quantum gravity in AdS. In this instance we have effectively identified a set of facts to import which uniquely pin down an amplitude.  These constraints are fairly natural in gravity -- they reflect the non-factorization of 3d gravity between the two boundaries -- but they are not the standard inputs used in the bootstrap community to study holographic CFTs, like those mentioned above.

We view this approach as complementary to our previous one~\cite{Cotler:2020ugk}. Our results here imply that a direct quantization of pure 3d gravity gives the same result as the ``constrain first'' approach. We infer that, after imposing the constraints in the direct quantization, the ghost path integral is simply ``1'', and one recovers the quantization of the classical phase space. It is also our hope that a more algebraic approach along the lines of the present paper will allow us to compute the gravity path integral on more complicated 3-manifolds.

The authors of~\cite{Afkhami-Jeddi:2020ezh,Maloney:2020nni} have recently considered an ensemble of $D$ free bosons on a Narain lattice where one averages over the moduli, which is closely related to an $\mathbb{R}^D\times \mathbb{R}^D$ Chern-Simons theory in three dimensions. It is instructive to compare and contrast the result for $\langle Z(\tau_1,\bar{\tau}_1),Z(\tau_2,\bar{\tau}_2)\rangle_{\text{ensemble, conn.}}$ in gravity and for the Narain ensemble. We do so for general $D$. We can think of the gravitational and Narain answers as giving two different solutions to a random CFT ``modular bootstrap'' problem, corresponding to very different spectral statistics. The wormhole amplitude in gravity encodes the physics of eigenvalue repulsion in the spectrum of BTZ microstates. There is an analogous ``wormhole amplitude'' in the fluctuation statistics of the Narain average, corresponding to Chern-Simons theory on the torus times interval. But this contribution instead encodes ultralocal attractive interactions between microstates, reflecting the underlying discreteness of the CFT spectrum, a feature already noted in~\cite{Maloney:2020nni}. We also study the decomposition of the Narain ensemble into symmetry sectors.  In a sentence, the 3d gravity results are consistent with a dual ensemble of CFTs with a chaotic spectrum of heavy states, while we find that the Narain ensemble comprises non-chaotic CFTs. 

At the end of the paper, we explain how our approach aligns with a broader perspective on bootstrapping random CFT. We sketch how such a bootstrap may be developed, and novel features it may have.

%------------------------------------------
\section{Modular bootstrap}
\label{S:bootstrap}
%------------------------------------------

\begin{figure}[t]
	\begin{center}
	\includegraphics[width=4in]{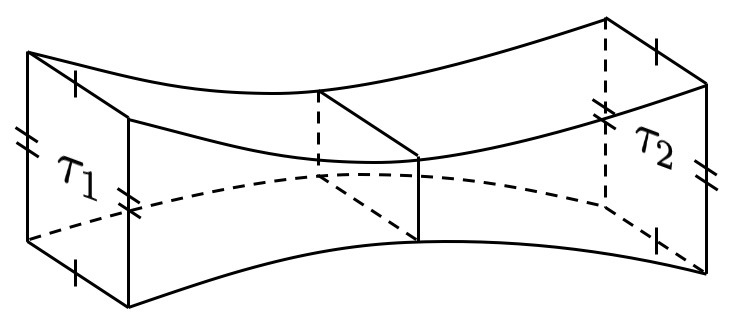}
	\end{center}
	\caption{\label{F:wormhole} The Euclidean wormhole $\mathbb{T}^2\times I$. For this configuration, the spatial and temporal circles on boundary $1$ continuously interpolate to the spatial and temporal circles on boundary $2$. Other configurations have a relative Dehn twist between the boundary tori.}
\end{figure}

The torus times interval amplitude in~\eqref{E:mainResult} may be written as
\beq
\label{E:fullZ}
	Z_{\mathbb{T}^2\times I}(\tau_1,\bar{\tau}_1,\tau_2,\bar{\tau}_2) = \sum_{\gamma\in PSL(2;\mathbb{Z})} \widetilde{Z}(\tau_1,\bar{\tau}_1,\gamma\tau_2,\gamma\bar{\tau}_2)\,,
\eeq
where the ``preamplitude'' $\widetilde{Z}$ is
\beq
\label{E:preAmplitude}
	\widetilde{Z}(\tau_1,\bar{\tau}_1,\tau_2,\bar{\tau}_2) = \frac{1}{2\pi^2}Z_0(\tau_1,\bar{\tau}_1)Z_0(\tau_2,\bar{\tau}_2) \frac{\text{Im}(\tau_1)\text{Im}(\tau_2)}{|\tau_1+\tau_2|^2}\,.
\eeq
The preamplitude has the interpretation of the gravitational path integral over the torus times interval, with no Dehn twist between the two torus boundaries. See Fig.~\ref{F:wormhole}. As such it is invariant under simultaneous modular transformations,
\beq
\label{E:modular}
	\widetilde{Z}(\gamma \tau_1,\gamma\bar{\tau}_1,\gamma^{-1}\tau_2,\gamma^{-1}\bar{\tau}_2) = \widetilde{Z}(\tau_1,\bar{\tau}_1,\tau_2,\bar{\tau}_2)\,.
\eeq
The full path integral is then a sum over configurations where one asymptotic region is twisted relative to the other. These Dehn twists are parameterized by elements of $PSL(2;\mathbb{Z})$, and lead to the modular sum in~\eqref{E:fullZ}.

In this Section we show that the preamplitude $\widetilde{Z}$ is completely fixed by consistency conditions after importing some basic facts from 3d gravity on the torus times interval. These facts are as follows:
\begin{enumerate}
	\item There are two twist zero modes, corresponding to spacetime translations along one boundary relative to another. The zero mode volume is~\cite{Cotler:2020ugk}, after a choice of normalization,
	\beq
	\label{E:zeroMode}
		V_0 = \sqrt{\text{Im}(\tau_1)\text{Im}(\tau_2)}\,.
	\eeq
	\item Each component of the boundary hosts two copies of a Virasoro symmetry. As a result, each boundary component is weighted by the character of a non-degenerate representation of the Virasoro group with some scaling weights $h,\bar{h}$.
	\item The two boundaries perceive the same energy and angular momentum, i.e.~the scaling weights $(h_1,\bar{h}_1)$ on boundary 1 equal those on boundary 2, $h_1=h_2=h$ and $\bar{h}_1=\bar{h}_2=\bar{h}$. 
	\item The scaling weights $h$ and $\bar{h}$ obey $h,\bar{h}\geq \frac{c-1}{24}$ with $c$ the exact central charge. This bound is the na\"{i}ve threshold for BTZ black hole production in pure 3d gravity coming from the Maloney-Witten density of states~\cite{Maloney:2007ud,Keller:2014xba}.\footnote{Another way of arriving at this bound is the following. Each boundary is equipped with two chiral edge modes~\cite{Cotler:2018zff,Cotler:2020ugk} which are analogues of the Schwarzian mode~\cite{Jensen:2016pah,Maldacena:2016upp,Engelsoy:2016xyb} in Jackiw-Teitelboim gravity. These edge modes produce a one-loop exact renormalization of the central charge, $c = \frac{3}{2G}+1$, and the path integral over them produces the character of a non-degenerate representation of the Virasoro group with scaling weights $h ,\bar{h}$. Non-singularity of the bulk geometry implies $h,\bar{h} \geq \frac{C}{24}$ with $C = \frac{3}{2G}$ the bare central charge. Conversely, the bulk geometry develops a conical singularity if $h$ or $\bar{h}$ is less than $\frac{C}{24}$.} The scaling weights may be parameterized in terms of ``momenta'' $k$ and $\bar{k}$ as
	\beq
		h -\frac{c-1}{24}= \frac{k^2}{4}\,, \qquad \bar{h} -\frac{c-1}{24}= \frac{\bar{k}^2}{4}\,.
	\eeq
\end{enumerate}
The second and fourth facts follow from the standard holographic dictionary for AdS$_3$/CFT$_2$, whereas the first and third facts are more novel. For $\text{Im}(\tau_1)=\text{Im}(\tau_2)$ so that $V_0 \propto \text{Im}(\tau)$, the first fact can be anticipated as a statement that the zero mode volume of the relative translation is proportional to the volume of the thermal circle, but for $\text{Im}(\tau_1)\neq \text{Im}(\tau_2)$ it is a non-trivial input from gravity. The third fact holds because energy and angular momentum are subject to Gauss' law constraints that equate them on the two boundaries in the absence of bulk matter (see~\cite{Cotler:2020ugk} for more details).

The most general $\widetilde{Z}$ consistent with the four facts above depends on a single function $\rho(k,\bar{k})$ where without loss of generality we take $k,\bar{k}\geq 0$. Specifically, $\widetilde{Z}$ may be decomposed as
\beq
\label{E:ansatz}
	\widetilde{Z}(\tau_1,\bar{\tau}_1,\tau_2,\bar{\tau}_2) = V_0 \int_0^{\infty} dkd\bar{k} \,\chi_k(\tau_1)\chi_k(\tau_2)  \overline{\chi}_{\bar{k}}(\bar{\tau}_1)\overline{\chi}_{\bar{k}}(\bar{\tau}_2)\, \rho(k,\bar{k})\,.
\eeq
Here $\chi_k(\tau)$ is a non-degenerate holomorphic Virasoro character,
\beq
\chi_k(\tau) = \frac{q^{\frac{k^2}{4}}}{\eta(\tau)}\,,
\eeq
and $\overline{\chi}_{\bar{k}}(\bar{\tau})$ a non-degenerate anti-holomorphic Virasoro character.

It remains to impose modular invariance~\eqref{E:modular}. The ansatz~\eqref{E:ansatz} is already invariant under a $T$-transformation, which acts as $\tau_1\to \tau_1+1$ and $\tau_2\to \tau_2-1$. So we need to impose invariance under an $S$-transformation, which after using $\text{Im}\left(-\frac{1}{\tau}\right) = \frac{\text{Im}(\tau)}{|\tau|^2}$ implies
\begin{align}
\begin{split}
	\int_0^{\infty}dkd\bar{k}\, \frac{\chi_k\left( -\frac{1}{\tau_1}\right)}{\sqrt{-i\tau_1}} \frac{\chi_k\left( -\frac{1}{\tau_2}\right)}{\sqrt{-i\tau_2}}&\frac{\overline{\chi}_{\bar{k}}\left(-\frac{1}{\bar{\tau}_1}\right)}{\sqrt{i\bar{\tau}_1}} \frac{\overline{\chi}_{\bar{k}}\left(-\frac{1}{\bar{\tau}_2}\right)}{\sqrt{i\bar{\tau}_2}}\,\rho(k,\bar{k}) \\
	& = \int_0^{\infty}dkd\bar{k} \,\chi_k(\tau_1)\chi_k(\tau_2)\overline{\chi}_{\bar{k}}(\bar{\tau}_1)\overline{\chi}_{\bar{k}}(\bar{\tau}_2)\,\rho(k,\bar{k})\,.
\end{split}
\end{align}
We will utilize a crossing kernel $S_{kk'}$ that allows us to express $\frac{\chi_k(-1/\tau)}{\sqrt{i \tau}}$ in terms of ordinary characters $\chi_{k'}(\tau)$,
\beq
\label{E:crossing}
	\frac{\chi_k\left(-\frac{1}{\tau}\right)}{\sqrt{-i\tau}} =\int_0^{\infty} dk'\,S_{kk'} \chi_{k'}(\tau)\,, \qquad S_{kk'} = \pi k' J_0(\pi k k')\,,
\eeq
which we derive at the end of this Section.  Using \eqref{E:crossing}, we see that
\begin{align}
\begin{split}
	\int_0^{\infty} dk d\bar{k}dk'dk''d\bar{k}'d\bar{k}''\, \chi_{k'}(\tau_1)\chi_{k''}(\tau_2)& \overline{\chi}_{\bar{k}'}(\bar{\tau}_1)\overline{\chi}_{\bar{k}''}(\bar{\tau}_2) \,S_{kk'}S_{kk''}S_{\bar{k}\bar{k}'}S_{\bar{k}\bar{k}''}\rho(k,\bar{k})
	\\
	& = \int_0^{\infty}dk d\bar{k}\,\chi_k(\tau_1)\chi_k(\tau_2)\overline{\chi}_{\bar{k}}(\bar{\tau}_1)\overline{\chi}_{\bar{k}}(\bar{\tau}_2)\,\rho(k,\bar{k})\,.
\end{split}
\end{align}
The non-degenerate Virasoro characters $\chi_k(\tau)$ provide a basis for functions of $\tau$ (that are at most as singular as $\frac{1}{\eta(\tau)}$ as $\tau\to i\infty$, as the above integrals are), and so it follows that 
\beq
\label{E:almostThere}
	\frac{\delta(k'-k'')}{k'}\frac{\delta(\bar{k}'-\bar{k}'')}{\bar{k}'} \frac{\rho(k',\bar{k}') }{k'\bar{k}'}=\pi^4 \int_0^{\infty} dkd\bar{k}\, k \bar{k}J_0(\pi kk')J_0(\pi kk'')J_0(\pi \bar{k}\bar{k}')J_0(\pi \bar{k}\bar{k}'')\frac{\rho(k,\bar{k})}{k \bar{k}}\,.
\eeq
Now multiply both sides by $k''J_0(\pi k'' q')$ and $\bar{k}''J_0(\pi \bar{k}''\bar{q}')$, where we take $q',\bar{q}'$ to be any non-negative values. Integrating over $k''\geq 0$ and $\bar{k}''\geq 0$ and employing the standard orthonormality relation for Bessel functions (here $k',k''>0$)
\beq
	\int_0^{\infty} dk\,k J_0(k k') J_0(k k'') = \frac{\delta(k'-k'')}{k'}\,,
\eeq
we obtain
\beq
	\frac{\rho(k',\bar{k}')}{k' \bar{k}'} = \frac{\rho(q',\bar{q}')}{q' \bar{q}'}\,, \qquad \forall\, k',\bar{k}',q',\bar{q}'\geq 0\,.
\eeq
This only holds if each side is a constant, which determines $\rho(k,\bar{k})$ up to a normalization $\mathcal{C}$,
\beq
	\rho(k,\bar{k}) = 8\mathcal{C} k \bar{k}\,.
\eeq

So modular invariance fixes the preamplitude~\eqref{E:ansatz} to be (after performing the integrals over $k$ and $\bar{k}$)
\beq
	\widetilde{Z}(\tau_1,\bar{\tau}_1,\tau_2,\bar{\tau}_2) = \frac{\mathcal{C}}{2\pi^2} Z_0(\tau_1,\bar{\tau}_1)Z_0(\tau_2,\bar{\tau}_2) \frac{\text{Im}(\tau_1)\text{Im}(\tau_2)}{|\tau_1+\tau_2|^2}\,,
\eeq
which is simply the gravitational result~\eqref{E:preAmplitude} times a proportionality constant $\mathcal{C}$. Now we demand consistency with the Jackiw-Teitelboim limit of~\cite{Ghosh:2019rcj} at low temperature and fixed spin. But, since the gravitational result was already consistent with that limit, we see that
\beq
	\mathcal{C}=1\,,
\eeq
and we land on precisely the result~\eqref{E:mainResult} we computed in~\cite{Cotler:2020ugk}.

We conclude this Section by deriving the crossing equation~\eqref{E:crossing}. First, we may simplify the claimed identity by using the modular property of the Dedekind eta function $\eta(-1/\tau) = \sqrt{-i\tau} \,\eta(\tau)$ to give
\beq
	\frac{e^{-\frac{\pi i k^2}{2\tau}}}{-i\tau} = \int_0^{\infty} dk'S_{kk'} e^{\frac{\pi i k^2\tau}{2}}\,.
\eeq
Rather than directly proving this identity, it is easier to prove its Fourier transform with respect to the real part of $\tau$. This proves the desired identity after taking an inverse Fourier transform. Writing $\tau = x+iy$, multiply both sides by $e^{\frac{\pi q^2 y}{2}}$ and Fourier transform with respect to $x$. On the right-hand-side we obtain (taking $q>0$)
\beq
	\int_{-\infty}^{\infty}  dx \,e^{-\frac{\pi i q^2 \tau}{2}}\int_0^{\infty} dk'S_{kk'} e^{\frac{\pi i k'^2\tau}{2}} = 4\int_0^{\infty} dk' \,\delta(q^2-k'^2) \pi k' J_0(\pi kk') = 2\pi J_0(\pi k q)\,.
\eeq
On the left-hand-side we find
\begin{align}
\begin{split}
	\int_{-\infty}^{\infty} dx\, e^{-\frac{\pi i q^2\tau}{2}}\frac{e^{-\frac{\pi i k^2}{2\tau}}}{-i \tau}&=\int_{-\infty}^{\infty} dx\, e^{-\frac{\pi i q^2\tau }{2}}\sum_{n=0}^{\infty} \frac{\left(-\frac{\pi k^2}{2}\right)^n}{n!}\frac{1}{(y-i x)^{n+1}}
	\\
	& = 2\pi \sum_{n=0}^{\infty} \frac{\left( -\frac{(\pi kq)^2}{4}\right)^n}{n!^{2}} = 2\pi J_0(\pi k q)\,,
\end{split}
\end{align}
where in going from the first line to the second we have exchanged summation with integration, and performed the integral over $x$ by the residue theorem. So the Fourier transform of~\eqref{E:crossing} holds true, which establishes the desired crossing equation.

%------------------------------------------
\section{Universality in random CFT?}
\label{S:universality}
%------------------------------------------

The gravitational inputs described near Eq.~\eqref{E:zeroMode} were essential to arrive at the answer in~\eqref{E:preAmplitude}. As a thought exercise, suppose that we replace the zero mode volume Eq.~\eqref{E:zeroMode} with $V_0 =1$. We will see the significance of this choice shortly. Let us denote the density of states and preamplitude by $\rho'$ and $\widetilde{Z}'$. Then invariance under simultaneous modular transformations $\tau_1\to\gamma \tau_1,\tau_2\to\gamma^{-1}\tau_2$ again fixes the result for the preamplitude, which in this case reads
\beq
\label{E:otherPre}
	\widetilde{Z}'(\tau_1,\bar{\tau}_1,\tau_2,\bar{\tau}_2) = \mathcal{C}' \,Z_0(\tau_1,\bar{\tau}_1)Z_0(\tau_2,\bar{\tau}_2) \sqrt{\frac{\text{Im}(\tau_1)\text{Im}(\tau_2)}{|\tau_1+\tau_2|^2}} = \frac{1}{|\eta(\tau_1)|^2|\eta(\tau_2)|^2}\frac{\mathcal{C}'}{|\tau_1+\tau_2|}\,,
\eeq
corresponding to a constant density of states $ \rho'(k,\bar{k}) = 2\mathcal{C}'$ for some proportionality constant $\mathcal{C}'$. The modular sum $\sum_{\gamma\in PSL(2;\mathbb{Z})}\widetilde{Z}'(\tau_1,\bar{\tau}_1,\gamma\tau_2,\gamma\bar{\tau}_2)$ does not converge, but it can be easily regularized in a modular-invariant way.

Notice that, after stripping off the modular invariant product of $Z_0$'s, the preamplitude~\eqref{E:otherPre} is proportional to the square root of the gravitational result~\eqref{E:preAmplitude}. In fact, since the ratio $\frac{\text{Im}(\tau_1)\text{Im}(\tau_2)}{|\tau_1+\tau_2|^2}$ is invariant under simultaneous modular transformations, we can easily write down a candidate preamplitude in terms of an arbitrary function $\mathcal{G}$ as
\beq
	\widetilde{Z}_{\rm candidate}(\tau_1,\bar{\tau}_1,\tau_2,\bar{\tau}_2) = Z_0(\tau_1,\bar{\tau}_1)Z_0(\tau_2,\bar{\tau}_2) \,\mathcal{G}\left( \frac{\text{Im}(\tau_1)\text{Im}(\tau_2)}{|\tau_1+\tau_2|^2}\right)\,,
\eeq
where the gravitational result is $\mathcal{G}(x) =\frac{x}{2\pi^2}$ and the thought exercise above corresponds to $\mathcal{G}(x) = \mathcal{C}' \sqrt{x}$. 

Of course we did not invent this thought exercise out of the ether. It is the result one obtains for the amplitude of a cousin of 3d gravity on the ``wormhole'' geometry $\mathbb{T}^2\times I$. This cousin is $SO(2,2)$ Chern-Simons theory with appropriate boundary conditions, which is classically equivalent to 3d gravity on the disk times time, but which differs on more complex topologies like $\mathbb{T}^2\times I$ (see~\cite{Cotler:2020ugk} for more details). These boundary conditions are the usual ones in AdS$_3$ gravity, translated to Chern-Simons variables. On the torus times interval, equivalently the annulus times a circle, the $SO(2,2)$ Chern-Simons theory with these boundary conditions becomes two copies of Liouville theory. 

One can see this from~\eqref{E:otherPre} in the following way. The orientation on the two boundaries is the one obtained using an outward pointing normal vector field. Flipping the orientation on boundary $2$, we would have arrived at the amplitude~\eqref{E:otherPre} with the substitution $\tau_2 \to -\bar{\tau}_2$. With this choice, the left-movers for one of the Liouville copies live on boundary $1$ and the right-movers on boundary $2$, and the situation is flipped for the other Liouville copy. The constant $\mathcal{C}'$ should then be interpreted as the (infinite) field range of the Liouville zero modes. One can imagine redefining the problem to remove these zero modes by hand, leading to a finite answer. Ignoring this field range, the constant density of states can then be understood as the usual statement that Liouville theory has a constant density of scalar primaries.

So we see that a phenomenologically informed modular bootstrap is able to obtain wormhole amplitudes in two examples: 3d gravity, and its cousin $SO(2,2)$ Chern-Simons theory. 

In fact Eq.~\eqref{E:otherPre} is the same result we obtain for the torus times interval amplitude of yet another model, $\mathbb{R}\times \mathbb{R}$ Chern-Simons theory with an action $\sim \frac{1}{2\pi}\int a \wedge db$, after a similar deletion of zero modes. This Chern-Simons theory belongs to a family of models, namely $\mathbb{R}^D\times \mathbb{R}^D$ Chern-Simons theory with action $\frac{1}{2\pi} \int \sum_i a_i \wedge db_i$, which is in a sense a dual to an ensemble of 2d CFTs given by $D$ bosons on a Narain lattice~\cite{Afkhami-Jeddi:2020ezh,Maloney:2020nni}. (More precisely, the average partition function on a surface $\Sigma$ for a suitable distribution of lattice moduli is equal to a sum over Chern-Simons path integrals on handlebodies.) For general $D$ the wormhole amplitude is simply the result~\eqref{E:otherPre} raised to the $D$th power (after a suitable deletion of zero modes), giving
\beq
\label{E:narainWormhole}
	\widetilde{Z}'(\tau_1,\bar{\tau}_1,\tau_2,\bar{\tau}_2) = Z_0(\tau_1,\bar{\tau}_1)^DZ_0(\tau_2,\bar{\tau}_2)^D \left( \frac{\text{Im}(\tau_1)\text{Im}(\tau_2)}{|\tau_1+\tau_2|^2}\right)^\frac{D}{2}\,.
\eeq
Up to normalization, it is the unique answer for a preamplitude consistent with a $U(1)^D_L\times U(1)^D_R$ Kac-Moody symmetry and simultaneous modular invariance of the form
\beq
	\widetilde{Z}'(\tau_1,\bar{\tau}_1,\tau_2,\bar{\tau}_2) =\left( \int_0^{\infty} dk d\bar{k} \,\chi_k(\tau_1)\chi_k(\tau_2)\overline{\chi}_{\bar{k}}(\bar{\tau}_1)\overline{\chi}_{\bar{k}}(\bar{\tau_2})\,\rho(k,\bar{k})\right)^D\,,
\eeq
with $\rho(k,\bar{k}) = 2$. 

These wormhole contributions encode rather different physics in 3d gravity and in the Narain ensemble, both of which may be discussed in terms of the spectral form factor~\cite{brezin1997spectral, prange1997spectral, cotler2017black}. In 3d gravity~\cite{Cotler:2020ugk} we showed that the amplitude~\eqref{E:preAmplitude} informs us that BTZ black hole microstates near threshold repel each other, leading to a ramp in the spectral form factor of the putative dual. By contrast, in the Narain ensemble the wormhole amplitude encodes the fundamental discreteness of the CFT spectrum by contributing to a plateau in the spectral form factor,  as noted in~\cite{Maloney:2020nni}.

In the remainder of this Section we study the wormhole contributions to the Narain ensemble in some detail at low temperature. We then compare and contrast the 3d gravity result with those for the Narain ensemble. The former is what we would expect for an ensemble of CFTs with a chaotic spectrum of heavy states, while the latter is an ensemble of non-chaotic CFTs.

%------------------------------------------
\subsection{The Narain ensemble}
%------------------------------------------

Consider the Narain average recently studied in~\cite{Afkhami-Jeddi:2020ezh,Maloney:2020nni}. Expressions for the average torus partition function appear in~\cite{Afkhami-Jeddi:2020ezh,Maloney:2020nni}, and its two-point function is in~\cite{Maloney:2020nni}. The average partition function may be expressed as a modular sum,
\beq
\label{E:narain1pt}
	\langle Z(\tau,\bar{\tau})\rangle_{\rm Narain} = Z_0(\tau,\bar{\tau})^D\sum_{\gamma \in PSL(2;\mathbb{Z})/\Gamma_{\infty}}\text{Im}(\gamma \tau)^{\frac{D}{2}}\,.
\eeq
Here $\gamma  = \begin{pmatrix} a & b \\ c & d \end{pmatrix}$ and $\Gamma_{\infty}$ is the subgroup of $PSL(2;\mathbb{Z})$ which fixes $i\infty$. The elements of $\Gamma_\infty$ are of the form $\begin{pmatrix} a & b \\ 0 & d \end{pmatrix}$. 

We may also consider the average of the partition function $Z_{\Sigma}(\Omega)$ of the free CFT on a genus-2 surface $\Sigma$ with $2\times 2$ period matrix $\Omega$. The average may be expressed as a sum over the $Sp(4;\mathbb{Z})$ modular group of a genus-2 surface,
\beq
	\langle Z_{\Sigma}(\Omega,\overline{\Omega})\rangle_{\rm Narain} = Z_{0,\Sigma}(\Omega,\overline{\Omega})^D \sum_{\gamma \in Sp(4;\mathbb{Z})/P}\text{det}(\text{Im}(\gamma \Omega))^{\frac{D}{2}}\,, \quad Z_{0,\Sigma}(\Omega,\overline{\Omega}) = \frac{1}{\sqrt{\text{det}(\text{Im}(\Omega))}|\text{det}'(\bar{\partial}_{\Sigma})|}\,.
\eeq 
Here $Z_{0,\Sigma}(\Omega,\overline{\Omega})$ is the partition function of a ``non-compact boson'' on $\Sigma$, with $\text{det}'(\bar{\partial}_{\Sigma})$ the determinant of $\bar{\partial}$ on $\Sigma$ omitting zero modes. Elements of $Sp(4;\mathbb{Z})$ may be parameterized in terms of $2\times 2$ blocks as $\begin{pmatrix} A & B \\ C & D \end{pmatrix}$ with
\beq
	A B^T = B A^T\,, \qquad C D^T = D C^T \,, \qquad A D^T - B C^T = I_2\,.
\eeq
$P$ is the Siegel parabolic subgroup of $Sp(4;\mathbb{Z})$, the analogue of $\Gamma_{\infty}$ for genus-2 surfaces, and it is parameterized by elements of the form $\begin{pmatrix} A & B \\ 0 & D\end{pmatrix}$. The action of $Sp(4;\mathbb{Z})$ on the period matrix $\Omega$ is given by $\gamma\Omega = (A\Omega + B)(C \Omega + D)^{-1}$.

Convergence of the modular sum over $PSL(2;\mathbb{Z})$ requires $D>2$, and convergence of the sum over $Sp(4;\mathbb{Z})$ requires $D>3$. Both the average torus and genus-2 partition functions are manifestly modular invariant.

The full two-point function of the torus partition function is in fact equal to the average of the genus-2 partition function in the special case where the initial period matrix is diagonal, $\Omega = \begin{pmatrix} \tau_1 & 0 \\ 0 & \tau_2\end{pmatrix}$, corresponding to a genus-2 surface obtained by gluing together two tori of complex structures $\tau_1$ and $\tau_2$ at a point. The two-point function may be written as
\beq
\label{E:narain2pt}
	\langle Z(\tau_1,\bar{\tau}_1)Z(\tau_2,\bar{\tau}_2)\rangle_{\rm Narain} = Z_0(\tau_1,\bar{\tau}_1)^DZ_0(\tau_2,\bar{\tau}_2)^D \sum_{\gamma \in Sp(4;\mathbb{Z})/P}\text{det}\left( \text{Im}(\gamma \Omega)\right)^{\frac{D}{2}}\,.
\eeq
Because it is a limiting value of the average genus-2 partition function, it is invariant under the full $Sp(4;\mathbb{Z})$ modular group, which includes as a subgroup invariance under independent modular transformations of $\tau_1$ and $\tau_2$\,.

%------------------------------------------
\subsubsection{Primary counting partition functions}
%------------------------------------------

All CFT states contribute to the one-point and two-point functions in~\eqref{E:narain1pt} and~\eqref{E:narain2pt}. It is worthwhile to consider the correlation functions of the partition function $Z^P(\tau,\bar{\tau})$ which counts the contributions from primary states alone. For the Narain CFTs it is related to the torus partition function as
\beq
	Z(\tau,\bar{\tau}) = \left( \prod_{n=1}^{\infty}\frac{1}{|1-q^n|^2}\right)^D Z^P(\tau,\bar{\tau}) = Z_0(\tau,\bar{\tau})^D \left( \sqrt{\text{Im}(\tau)}|q|^{\frac{1}{12}}\right)^D Z^P(\tau,\bar{\tau})\,.
\eeq
Its one-point function is
\beq
	\langle Z^P(\tau,\bar{\tau})\rangle_{\rm Narain} =\left( \frac{|q|^{-\frac{1}{12}}}{\sqrt{\text{Im}(\tau)}}\right)^D\sum_{\gamma\in PSL(2;\mathbb{Z})/\Gamma_{\infty}} \text{Im}(\gamma \tau)^{\frac{D}{2}} = |q|^{-\frac{D}{12}} \sum_{c,d,\,(c,d)=1} \frac{1}{|c\tau+d|^D}\,.
\eeq
Fourier transforming to fixed spin $s$ one has for $s\neq 0$
\beq
\label{E:narain1ptSpin}
	\left\langle \text{tr}\big( e^{-\beta H_s}\big)\right\rangle_{\rm Narain} =e^{\frac{\pi D}{6}\beta} \left( \sum_{c>0} \frac{c_c(s)}{c^D} \right) \frac{2\pi^{\frac{D}{2}}}{\Gamma\left( \frac{D}{2}\right)} \left( \frac{|s|}{\beta}\right)^{\frac{D-1}{2}}K_{\frac{1-D}{2}}(2\pi |s|\beta)\,,
\eeq
where $H_s$ is the CFT Hamiltonian at spin $s$ and $c_q(s) =\sum_{0\leq d\leq c-1,(c,d)=1} e^{-2\pi i\frac{ d}{q}s}$ is Ramanujan's sum. The sum over $c$ converges for $D>1$. For $s=0$ one has
\beq
	\left\langle \text{tr}\big( e^{-\beta H_0}\big)\right\rangle_{\rm Narain} = e^{\frac{\pi D}{6}\beta} \left( 1 + \left( \sum_{c>0} \frac{\phi(c)}{c^D}\right) \frac{\sqrt{\pi}\, \Gamma\left( \frac{D-1}{2}\right)}{\Gamma\left( \frac{D}{2}\right)} \beta^{1-D}\right)\,,
\eeq
where $\phi(c)$ is the Euler totient function. The sum converges for $D>2$.

The two-point function is
\beq
\label{E:narain2ptP}
	\langle Z^P(\tau_1,\bar{\tau}_2)Z^P(\tau_2,\bar{\tau}_2)\rangle_{\rm Narain} = \left(\frac{|q_1q_2|^{-\frac{1}{12}}}{\sqrt{\text{Im}(\tau_1)\text{Im}(\tau_2)}}\right)^D \sum_{\gamma \in Sp(4;\mathbb{Z})/P} \text{det}(\text{Im}(\gamma \Omega))^{\frac{D}{2}}\,.
\eeq
The $Sp(4;\mathbb{Z})$ sum simplifies at low temperature as we will see momentarily.

%------------------------------------------
\subsubsection{Low-temperature limit}
%------------------------------------------

We now explore some features of the low-temperature limit of the connected two-point function of $Z^P$.  In particular, we study how the ``wormholes'' in the Narain ensemble contribute to a plateau in the spectral form factor~\cite{Maloney:2020nni}, with the aim of comparing to the wormhole amplitude in 3d gravity.  For a more detailed study of the 
spectral form factor we refer the reader to~\cite{maloneySFF}.\footnote{We thank A. Maloney for useful discussions related to this Subsection.}

 Now consider the low-temperature limit. Parameterizing $\tau_1=x_1+i\beta_1,\tau_2=x_2+i \beta_2$, this limit corresponds to taking $\beta_1,\beta_2\to \infty$ while holding $x_1,x_2$ and the ratio $\frac{\beta_1}{\beta_2}$ fixed. For the one-point function, using
\beq
	\text{Im}(\gamma\tau) = \frac{\text{Im}(\tau)}{|c\tau+d|^2}\,,
\eeq
the terms in the modular sum behave rather differently. The identity transformation $\gamma = I_2$ contributes $\text{Im}(\gamma\tau)^{\frac{D}{2}} = \beta^{\frac{D}{2}}$, while all other terms with $c\neq 0$ are parametrically suppressed as $\text{Im}(\gamma \tau) = O(\beta^{-\frac{D}{2}})$. 

Similarly, for the two-point function, using
\beq
	\text{det}(\text{Im}(\gamma \Omega)) = \frac{\text{det}(\text{Im}(\Omega))}{|\text{det}(C \Omega + D)|^2}\,,
\eeq
we see that the low-temperature limit of this determinant falls into one of three classes: (i) For the identity transformation, $\text{det}(\text{Im}(\gamma \Omega))^{\frac{D}{2}} = (\beta_1\beta_2)^{\frac{D}{2}}$ is of $O(\beta^D)$. (ii) For elements of $Sp(4;\mathbb{Z})/P$ with $C\neq 0$ but $\text{det}(C)=0$, one finds\footnote{Here and through the end of this Subsection, we abuse terminology slightly by letting $D$ refer to the $2\times 2$ block inside of an $Sp(4;\mathbb{Z})$ element $\gamma$, and to the number of bosons. We hope that what we mean is clear by the context.}
\beq
\label{E:SpImportant}
	\text{det}(\text{Im}(\gamma \Omega))^{\frac{D}{2}} =  \left( \frac{\text{Im}(\tau_1) \text{Im}(\tau_2)}{|p \tau_1+q \tau_2+\text{det}(D)|^2}\right)^{\frac{D}{2}} \,, 
\eeq
with  $p = C_{11} D_{22}-C_{21} D_{12}  ,$ and $ q = C_{22} D_{11}-C_{12} D_{21} $, i.e.~the summand is of $O(\beta^0)$. (iii) Elements of $Sp(4;\mathbb{Z})/P$ with $\text{det}(C)\neq 0$ lead to contributions which are suppressed at low temperature as $\text{det}(\text{Im}(\gamma\Omega))^{\frac{D}{2}}= O(\beta^{-D})$.

What about the wormhole contributions we identified at the beginning of this Section in~\eqref{E:narainWormhole}? Their contribution to the connected two-point function of $Z$ is 
\beq
	\langle Z(\tau_1,\bar{\tau}_1)Z(\tau_2,\bar{\tau}_2)\rangle_{\rm Narain,conn.} \supset Z_0(\tau_1,\bar{\tau}_1)Z_0(\tau_2,\bar{\tau}_2) \sum_{\gamma\in PSL(2;\mathbb{Z})} \left( \frac{\text{Im}(\tau_1)\text{Im}(\gamma \tau_2)}{|\tau_1+\gamma \tau_2|^2}\right)^{\frac{D}{2}}\,.
\eeq
In a putative 3d dual, these terms arise from $\mathbb{R}^D\times \mathbb{R}^D$ Chern-Simons theory on the  torus times interval, after a deletion of zero modes and where the sum is over Dehn twists of one torus boundary relative to the other. The dominant wormhole contributions at low temperature are those with $\gamma \tau_2 = \tau_2 + n$ and $n\in \mathbb{Z}$. These correspond to the $p=q=1$ terms in the full $Sp(4;\mathbb{Z})$ sum. 

The connected two-point function of $Z^P$, namely
\begin{align}
	\langle Z^P(\tau_1,\bar{\tau}_1)Z^P(\tau_2&,\bar{\tau}_2)\rangle_{\text{Narain, conn.}} = \left( \frac{|q_1q_2|^{-\frac{1}{12}}}{\sqrt{\text{Im}(\tau_1)\text{Im}(\tau_2)}}\right)^D 
	\\
	\nonumber
	&\times \left( \sum_{\gamma \in Sp(4;\mathbb{Z})/P}\text{det}(\text{Im}(\gamma\Omega))^{\frac{D}{2}} - \sum_{\gamma_1,\gamma_2\in PSL(2;\mathbb{Z})/\Gamma_{\infty}}\left( \text{Im}(\gamma_1 \tau_1)\text{Im}(\gamma_2\tau_2)\right)^{\frac{D}{2}}\right)\,,
\end{align}
therefore simplifies at low temperature. The identity term in the $Sp(4;\mathbb{Z})$ sum cancels against the $\gamma_1=\gamma_2=I_2$ contribution in the double sum over $PSL(2;\mathbb{Z})$; the terms in the $Sp(4;\mathbb{Z})$ sum with $C\neq 0$ but $\text{det}(C)=0$ in~\eqref{E:SpImportant} with $q=0$ cancel against the terms in the double sum over $PSL(2;\mathbb{Z})$ with $\gamma_1\neq I_2$ but $\gamma_2 = I_2$; the terms with $C\neq 0$, $\text{det}(C)=0$, and $p=0$ cancel against the terms with $\gamma_1=I_2$ but $\gamma_2 \neq I_2$. The low-temperature limit is therefore dominated by terms with $C\neq 0,\text{det}(C)=0$, with both $p,q\neq 0$. Let us denote the set of such elements as $\mathcal{S}$. Then the leading low-temperature limit is given by
\begin{align}
\nonumber
	\langle Z^P(\tau_1,\bar{\tau}_1)Z^P(\tau_2,\bar{\tau}_2)\rangle_{\text{Narain, conn.}} &=\left( \frac{|q_1q_2|^{-\frac{1}{12}}}{\sqrt{\text{Im}(\tau_1)\text{Im}(\tau_2)}}\right)^D \left( \sum_{\gamma \in \mathcal{S}} \text{det}(\text{Im}(\gamma\Omega))^{\frac{D}{2}} + O(\beta^{-D/2})\right)
	\\
	&= |q_1q_2|^{-\frac{D}{12}} \left( \sum_{\gamma \in \mathcal{S}}|p \tau_1+q \tau_2+\text{det}(D)|^{-\frac{D}{2}} + O(\beta^{-D})\right)\,.
\end{align}

Suppose we consider the Fourier transform to fixed identical spins, $s_1=s_2=s \neq 0$. To leading order in the low-temperature expansion only terms with $p=q$ contribute. These terms include the $p=q=1$ wormhole contributions, and up to additive shifts in the $\tau$'s, they all have the same functional form.\footnote{Note that the wormhole contributions at $p=q=1$ dominate in the large $D$ limit.} As a result, for a constant $\mathcal{C}_s$ arising from the modular sum we find
\begin{align}
\nonumber
	\left\langle \text{tr}\big( e^{-\beta_1 H_s}\big)\text{tr}\big( e^{-\beta_2 H_s}\big)\right\rangle_{\text{Narain, conn.}} &=\mathcal{C}_s \, e^{\frac{\pi D}{6} (\beta_1+\beta_2)}\left(  \frac{2\pi^{\frac{D}{2}}}{\Gamma\left(\frac{D}{2}\right)} \left( \frac{|s|}{\beta_1+\beta_2}\right)^{\frac{D-1}{2}}K_{\frac{1-D}{2}}\left( 2\pi |s|(\beta_1+\beta_2)\right) +\cdots\right)
	\\
	\label{E:narainPlateau}
	& =\mathcal{C}_s \, e^{-E_{s}(\beta_1+\beta_2)}\frac{\sqrt{2\pi |s|}\pi^{\frac{D}{2}}}{\Gamma\left(\frac{D}{2}\right)}\left(\frac{ |s|}{\beta_1+\beta_2}\right)^{\frac{D-2}{2}}\left( 1 + O(\beta^{-1})\right)\,,
\end{align}
where the Narain threshold energy is $E_s = 2\pi \left( |s|-\frac{D}{12}\right)$. The dots in the first line refer to the leading low-temperature corrections, which are parametrically suppressed relative to the indicated term by a factor of $\beta^{-\frac{D}{2}}$. Inverse Laplace transforming we find that the pair correlations are, near threshold $E\gtrsim E_s$\,,
\beq
\label{E:attraction}
	\langle \rho^P_s(E_1)\rho^P_s(E_2)\rangle_{\text{Narain, conn.}} \approx \mathcal{C}_s \,\frac{(2\pi^2|s|)^{\frac{D-1}{2}}}{2\sqrt{\pi}\,\Gamma\left(\frac{D}{2}\right)\Gamma\left(\frac{D-2}{2}\right)}\,E1^{\frac{D}{2}-2}\,\delta(E_1-E_2)\,,
\eeq
where $\rho^P_s(E)$ is the density of primary states of spin $s$. Away from threshold, but ignoring the subleading terms in the first line of~\eqref{E:narainPlateau}, the correlations are still delta-localized. The pair correlations are approximately ultralocal and attractive.

The low-temperature results in~\eqref{E:narainPlateau} and~\eqref{E:attraction} can be understood in terms of a plateau in the spectral form factor. Setting
\beq
	\beta_1 = \beta + i T\,, \qquad \beta_2 = \beta - i T\,,
\eeq
so that the traces have the interpretation of an evolution operator at inverse temperature $\beta$ by a Lorentzian time $T$, we have
\begin{align}
\begin{split}
	&\left\langle \text{tr}\big( e^{-(\beta + i T)H_s}\big)\text{tr}\big( e^{-(\beta-i T)H_s}\big)\right\rangle_{\text{Narain, conn.}} 
	\\
	& \qquad \qquad \qquad \qquad \qquad = \mathcal{C}_s \,e^{\frac{\pi D}{3}\beta} \left( \frac{2\pi^{\frac{D}{2}}}{\Gamma\left( \frac{D}{2}\right)} \left(\frac{|s|}{2\beta}\right)^{\frac{D-1}{2}}K_{\frac{1-D}{2}}\left(4\pi |s|\beta\right) + O(\beta^{-D}e^{-4\pi |s|\beta})\right)\,,
\end{split}
\end{align}
which becomes a constant independent of time $T$ at late times. With some effort one can show that the corrections decay to zero in the $T\to\infty$ limit. This constant asymptote is the plateau in the spectral form factor of primaries at fixed spin. Comparing against the average value of $\text{tr}\left( e^{-\beta H_s}\right)$ in~\eqref{E:narain1ptSpin}, we see that the two-point function asymptotes to the same functional form as the one-point function at an inverse temperature $2\beta$, up to the proportionality constant. Indeed, one expects on general grounds for any ensemble of systems with a discrete spectrum that the two-point function approaches a constant at late times, the plateau, with
\beq
	\lim_{T\to\infty}\langle Z(\beta+iT)Z(\beta-iT)\rangle_{\text{ensemble, conn.}} = \langle Z(2\beta)\rangle_{\rm ensemble}\,.
\eeq
The analysis of~\cite{maloneySFF} demonstrates that this is exactly the case for the Narain ensemble.

Let us compare these results with 3d gravity. For gravity, inside the connected two-point function so that the vacuum contribution drops out, $Z^P$ is related to the ordinary partition function by \beq
	Z(\tau,\bar{\tau}) = \left( \prod_{n=1}^{\infty}\frac{1}{|1-q^n|^2}\right) Z^P(\tau,\bar{\tau}) = Z_0(\tau,\bar{\tau})\sqrt{\text{Im}(\tau)}|q|^{\frac{1}{12}}Z^P(\tau,\bar{\tau})\,.
\eeq
From the torus times interval amplitude we then have
\beq
	\langle Z^P(\tau_1,\bar{\tau}_1)Z^P(\tau_2,\bar{\tau}_2)\rangle_{\text{ensemble, conn.}} = \frac{1}{2\pi^2\sqrt{\text{Im}(\tau_1)\text{Im}(\tau_2)}}\,|q_1q_2|^{-\frac{1}{12}}\!\sum_{\gamma \in PSL(2;\mathbb{Z})} \frac{\text{Im}(\tau_1)\text{Im}(\gamma \tau_2)}{|\tau_1+\gamma\tau_2|^2} + \cdots\,,
\eeq
where the dots come from other connected topologies. At low temperature the modular sum is dominated by translations $\gamma \tau_2 = \tau_2 + n$ for $n\in \mathbb{Z}$. Fourier transforming to fixed spin we then found
\beq
	\left\langle \text{tr}\big( e^{-\beta_1H_{s_1}}\big)\text{tr}\big(e^{-\beta_2H_{s_2}}\big)\right\rangle_{\text{ensemble, conn.}} = e^{-E_{s_1}\beta_1-E_{s_2}\beta_2}\left(\frac{1}{2\pi}\frac{\sqrt{\beta_1\beta_2}}{\beta_1+\beta_2}  \,\delta_{s_1s_2} + O(\beta^{-1})\right) + \cdots\,,
\eeq
where $E_s = 2\pi \left( |s|-\frac{1}{12}\right)$ is the threshold energy to produce a BTZ black hole of spin $s$. The indicated term dominates the torus times interval amplitude at low temperature. As we discussed in~\cite{Cotler:2020ugk} this is precisely the result predicted by a random matrix theory ansatz.

Inverse Legendre transforming to fixed energy, one finds that for energies near threshold, i.e.~near $E_s$\,, the pair correlation of energy eigenvalues at fixed spin is that of double-scaled random matrix theory for large Hermitian matrices, 
\beq
\label{E:repulsion}
	\langle \rho^P_{s_1}(E_1)\rho^P_{s_2}(E_2)\rangle_{\text{ensemble, conn.}} \approx -\frac{1}{(2\pi)^2}\frac{E_1-E_{s_1}+E_2-E_{s_2}}{\sqrt{E_1-E_{s_1}}\sqrt{E_2-E_{s_2}}(E_1-E_2)^2} \,\delta_{s_1s_2}\,, 
\eeq
where $E_i>E_{s_i}$. In other words, energy eigenstates at fixed spin repel. 

%------------------------------------------
\subsection{Comparison and comments}
%------------------------------------------

As we have shown, a well-informed modular bootstrap can pinpoint the torus times interval amplitude of 3d gravity, as well of the Chern-Simons-like dual to the Narain ensemble. The fluctuation statistics encoded in these wormhole amplitudes are rather different. The 3d gravity amplitude describes the physics of energy eigenvalue repulsion in the spectrum of BTZ microstates, while the ``Narain wormholes'' contribute to a part of the plateau in the spectral form factor of the Narain ensemble at fixed spin (as observed in~\cite{Maloney:2020nni})  In fact in the large $D$ limit, the closest thing to a semiclassical approximation for the Narain ensemble, these wormholes fully account for the plateau to leading order in $1/D$.

Let us compare these examples a bit further, in the hope of gleaning some general lessons about CFT ensembles. 

In the Narain ensemble wormholes contribute to the plateau in the spectral form factor. One might wonder if there is also a ramp, i.e.~if there is an analogue of eigenvalue repulsion visible at earlier Lorentzian time. Besides processing the full $Sp(4;\mathbb{Z})$ sum in~\eqref{E:narain2ptP} (see~\cite{maloneySFF} for a detailed analysis beyond the scope of wormholes), we can get further information by studying Narain CFTs at fixed values of the moduli. In particular, after simultaneously diagonalizing the Hamiltonian, momentum, and other local symmetry charges, we can look at the nearest-neighbor statistics of energy eigenstates within a symmetry block. However in a Narain CFT, after accounting for the Kac-Moody symmetry, there is only a single energy eigenvalue per block. 

To see this recall the expression for the right- and left-moving energy eigenvalues. These are labeled by momentum and winding vectors $n_i$ and $w^i$ with $i=1,2,...,D$. The eigenvalues also depend on the target space metric $G_{ij}$ and NS $B$-field $B_{ij}$. They are given by (in $\alpha'=1$ units)
\beq
	L_0 = \frac{1}{4}G^{ij}(v_i + w_i)(v_j+w_j)\,, \qquad \overline{L}_0 = \frac{1}{4}G^{ij}(v_i-w_i)(v_i - w_i)\,, \qquad v_i = n_i + B_{ij}w^j\,,
\eeq
where $w_i = G_{ij} w^j$. In particular, the spin is given by $s = n_i w^i$. On account of the Kac-Moody symmetry, the expressions for $L_0$ and $\overline{L}_0$ are redundant. They are determined by the eigenspectrum of $D$ right-moving charges $Q_a$ and $D$ left-moving charges $\overline{Q}_a$, with
\beq
	L_0 = \frac{Q_a Q_a}{4}\,, \qquad \overline{L}_0 = \frac{\overline{Q}_a \overline{Q}_a}{4}\,.
\eeq 
To write the spectra of the Kac-Moody charges it is convenient to decompose the target space metric $G_{ij}$ into a coframe as $G_{ij} = \delta_{ab}e^a_i e^b_j$. Then
\beq
	Q_a = e_a^i(v_i+w_i)\,, \qquad \overline{Q}_a = e_a^i (v_i - w_i)\,.
\eeq
Inverting these, it follows that there is at most a single state at fixed values of the charges $Q_a$ and $\overline{Q}_b$. 

So, while irrational Narain CFTs (which compose most of the ensemble) have an erratic, infinite spectrum of primary states, they are ultimately non-chaotic in the sense above. After all, there is no eigenvalue repulsion when a symmetry block has a single eigenvalue.

By contrast, 3d gravity only has ordinary Virasoro symmetry, and so symmetry blocks are subspaces of fixed spin. There is an enormous $O(e^c)$ density of primary states at fixed spin, with $c$ the central charge, and our results in~\cite{Cotler:2020ugk} indicate that there is eigenvalue repulsion within each symmetry block.  These features are characteristic of a chaotic spectrum.

More broadly, we expect that irrational $c\gg 1$ CFTs without extended chiral algebras (assuming such CFTs exist) have chaotic spectra, including level repulsion within each symmetry block.  A typical such $c\gg 1$ CFT with sparse spectrum should have a very dense set of CFT microstates in the Cardy regime $h,\bar{h} \geq \frac{c-1}{24}$.  These statements would also hold in a statistical sense for an ensemble average over irrational $c\gg 1$ CFTs, and in the Cardy regime we expect that the leading eigenvalue fluctuation statistics are well-approximated by our gravity wormhole result.  This leads us to conjecture that the 3d gravity wormhole amplitude near threshold is the random CFT analogue of the universality of the double resolvent in double-scaled random matrix theory~\cite{eynard2007invariants}.

%------------------------------------------
\section{Towards bootstrapping CFT ensembles}
%------------------------------------------

In Sections~\ref{S:bootstrap} and~\ref{S:universality} we used a type of modular bootstrap to determine the torus times interval amplitude for AdS$_3$ gravity and for the ``gravity dual'' to the average over Narain CFTs. To land on these amplitudes we had to use a number of inputs from three dimensions. In our gravity analysis perhaps the most important of these were the zero mode volume $V_0$ and the constraint that scaling weights for operators on the two boundaries were linked. This modular bootstrap differs, both technically and conceptually, from existing approaches to the conformal bootstrap as we now explain.

In the usual bootstrap one can constrain the torus partition function of a single CFT. If one wanted to, one could instead constrain
\beq
\label{E:ZZ1}
	Z(\tau_1, \bar{\tau}_1) Z(\tau_2, \bar{\tau}_2)\,,
\eeq
the product of torus partition functions with different modular parameters $\tau_1$ and $\tau_2$.  However this would be tantamount to constraining each copy individually since~\eqref{E:ZZ1} tautologically factorizes.

However, we could instead attempt bootstrapping a putative ensemble of local CFTs at the same central charge (with the optimistic disposition that there are a landscape of such models at large central charge). We would like to constrain more complicated averages like
\beq
\label{E:ZZens1}
	\langle Z(\tau_1, \bar{\tau}_1) Z(\tau_2, \bar{\tau}_2) \rangle_{\text{ensemble}}\,.
\eeq
This quantity may not be reconstructed from $\langle Z(\tau, \bar{\tau})\rangle_{\text{ensemble}}$ on account of ensemble correlations. The lack of factorization has no counterpart in the standard CFT bootstrap.  In this paper, we have successfully bootstrapped (contributions to) the connected part of~\eqref{E:ZZens1} in two cases where we had strong constraints.  More generally, one could develop bootstrap techniques for~\eqref{E:ZZens1}, and specifically its connected part, instead imposing more traditional requirements like a large twist gap, large central charge, etc.

What are the ``observables'' that may be bootstrapped? These would be partition functions and any operators whose quantum numbers are constant across the distribution, including at least the stress tensor. The ensemble average of a primary operator is, in general, not a primary of the average.\footnote{As a simple example consider averaging over two instances of a compact boson at different values of the radius.} This is perhaps a feature rather than a bug when it comes to making contact with pure models of quantum gravity like 3d gravity or minimal AdS$_3$ supergravity.

A bootstrap for the average torus partition function $\langle Z(\tau,\bar{\tau})\rangle_{\rm ensemble}$ would essentially be the modular bootstrap for a non-compact CFT. One would decompose the average partition function into a sum of Virasoro characters convolved with the average density of states $\langle \rho(h,\bar{h})\rangle_{\rm ensemble}$\,. One would then impose modular invariance of the average partition function and a non-negative average density of states. For an ensemble of local CFTs without an extended chiral algebra, $\langle \rho(h,\bar{h})\rangle_{\rm ensemble}$ will contain a delta-localized piece corresponding to the vacuum representation, perhaps other delta-localized pieces (like the light operators of the two-dimensional supersymmetric SYK model~\cite{murugan2017more}) as well as a continuous spectrum.  

More generally one would like to bootstrap $n$-point averages of the form
\beq
\label{E:generalcorr}
	\langle Z_{\Sigma_{g_1}}\!(\Omega_1, \overline{\Omega}_1)  \cdots Z_{\Sigma_{g_n}}\!(\Omega_n, \overline{\Omega}_n) \rangle_{\text{ensemble}}
\eeq
where $Z_{\Sigma_g}(\Omega, \overline{\Omega})$ is the partition function on a genus $g$ surface with period matrix $\Omega$.  Since we are considering ensembles, an interesting feature is that $k$-point averages constrain $n$-point averages for $k<n$.

A simple analogy is instructive.  We can view~\eqref{E:generalcorr} as an $n$th moment of a CFT ensemble.  Given a probability density function $p(x)$, there are positivity constraints on its moments; the simplest nontrivial one is
\begin{equation}
\langle x^2 \rangle - \langle x \rangle^2 \geq 0\,.
\end{equation}
This constrains the second moment $\langle x^2 \rangle$ in terms of the first moment $\langle x \rangle$.  Many other such relations can be derived.  For CFT ensembles the simplest bound is that $\langle \rho(h_1,\bar{h}_1)\rho(h_2,\bar{h}_2)\rangle_{\text{ensemble, conn.}}$ is positive semi-definite as a kernel.

Beyond bounds one would like to have Ward identities which relate the different $n$-point averages to each other.  While it is not clear if such Ward identities exist for ensembles of 2d CFTs, they do exist in random matrix theory. Recently, a bootstrap approach was studied in the context of random matrix ensembles \cite{lin2020bootstraps}; there the Ward identities are packaged as loop equations.  (For a different approach to bootstrapping CFT ensembles, we refer the reader to \cite{komargodski2017random} which analyzes the random bond Ising model.)

An interesting possibility is suggested by the logic of statistical mechanics (or more properly, Bayesian inference).  Suppose we have a probability distribution $p(x)$ for which its first two moments are $\langle x \rangle = a$ and $\langle x^2 \rangle = b$.  Then the maximum entropy probability distribution consistent with these moments is a Gaussian.  As a consequence, all higher moments of the Gaussian are determined by the first two moments.  More generally, perhaps ``maximum entropy'' ansatzae are useful for bootstrapping CFT ensemble.  There is precedence for this in random matrix theories \cite{eynard2015random, Saad:2019lba, stanford2019jt}.  Remarkably, using topological recursion \cite{eynard2007invariants} (see \cite{eynard2015random} for a review), double-scaled random matrix theories are completely determined by their leading order contributions to the resolvent, and connected double resolvent.

As a final remark, it is possible that bootstrapping CFT ensembles may be easier than bootstrapping individual CFT's. Our investigations suggest that this may be the case for pure AdS$_3$ gravity. Furthermore, intuition about ensembles of chaotic CFTs can bring additional constraints to bear, such as having Virasoro random matrix nearest-neighbor energy level statistics \cite{Cotler:2020ugk}.

\subsection*{Acknowledgements}

We would like to thank Anatoly Dymarsky, Nicholas Hunter-Jones, Alexander Maloney and Edward Witten for valuable discussions. JC is supported by a Junior Fellowship from the Harvard Society of Fellows. KJ is supported in part by the Department of Energy under grant number DE-SC 0013682.

\bibliographystyle{JHEP}
\bibliography{refs}

\providecommand{\href}[2]{#2}\begingroup\raggedright\begin{thebibliography}{10}

\bibitem{Cotler:2020ugk}
J.~Cotler and K.~Jensen, {\it {AdS$_3$ gravity and random CFT}},
  \href{http://xxx.lanl.gov/abs/2006.08648}{{\tt 2006.08648}}.

\bibitem{Maldacena:2004rf}
J.~M. Maldacena and L.~Maoz, {\it {Wormholes in AdS}},  {\em JHEP} {\bf 02}
  (2004) 053, [\href{http://xxx.lanl.gov/abs/hep-th/0401024}{{\tt
  hep-th/0401024}}].

\bibitem{Saad:2019lba}
P.~Saad, S.~H. Shenker, and D.~Stanford, {\it {JT gravity as a matrix
  integral}},  \href{http://xxx.lanl.gov/abs/1903.11115}{{\tt 1903.11115}}.

\bibitem{Afkhami-Jeddi:2020ezh}
N.~Afkhami-Jeddi, H.~Cohn, T.~Hartman, and A.~Tajdini, {\it {Free partition
  functions and an averaged holographic duality}},
  \href{http://xxx.lanl.gov/abs/2006.04839}{{\tt 2006.04839}}.

\bibitem{Maloney:2020nni}
A.~Maloney and E.~Witten, {\it {Averaging Over Narain Moduli Space}},
  \href{http://xxx.lanl.gov/abs/2006.04855}{{\tt 2006.04855}}.

\bibitem{Belin:2020hea}
A.~Belin and J.~de~Boer, {\it {Random Statistics of OPE Coefficients and
  Euclidean Wormholes}},  \href{http://xxx.lanl.gov/abs/2006.05499}{{\tt
  2006.05499}}.

\bibitem{Maxfield:2020ale}
H.~Maxfield and G.~J. Turiaci, {\it {The path integral of 3D gravity near
  extremality; or, JT gravity with defects as a matrix integral}},
  \href{http://xxx.lanl.gov/abs/2006.11317}{{\tt 2006.11317}}.

\bibitem{Ghosh:2019rcj}
A.~Ghosh, H.~Maxfield, and G.~J. Turiaci, {\it {A universal Schwarzian sector
  in two-dimensional conformal field theories}},  {\em JHEP} {\bf 05} (2020)
  104, [\href{http://xxx.lanl.gov/abs/1912.07654}{{\tt 1912.07654}}].

\bibitem{Elitzur:1989nr}
S.~Elitzur, G.~W. Moore, A.~Schwimmer, and N.~Seiberg, {\it {Remarks on the
  Canonical Quantization of the Chern-Simons-Witten Theory}},  {\em Nucl. Phys.
  B} {\bf 326} (1989) 108--134.

\bibitem{Axelrod:1989xt}
S.~Axelrod, S.~Della~Pietra, and E.~Witten, {\it {Geometric quantization of
  Chern-Simons gauge theory}},  {\em J. Diff. Geom.} {\bf 33} (1991), no.~3
  787--902.

\bibitem{Maloney:2007ud}
A.~Maloney and E.~Witten, {\it {Quantum Gravity Partition Functions in Three
  Dimensions}},  {\em JHEP} {\bf 02} (2010) 029,
  [\href{http://xxx.lanl.gov/abs/0712.0155}{{\tt 0712.0155}}].

\bibitem{Keller:2014xba}
C.~A. Keller and A.~Maloney, {\it {Poincare Series, 3D Gravity and CFT
  Spectroscopy}},  {\em JHEP} {\bf 02} (2015) 080,
  [\href{http://xxx.lanl.gov/abs/1407.6008}{{\tt 1407.6008}}].

\bibitem{Cotler:2018zff}
J.~Cotler and K.~Jensen, {\it {A theory of reparameterizations for AdS$_3$
  gravity}},  {\em JHEP} {\bf 02} (2019) 079,
  [\href{http://xxx.lanl.gov/abs/1808.03263}{{\tt 1808.03263}}].

\bibitem{Jensen:2016pah}
K.~Jensen, {\it {Chaos in AdS$_2$ Holography}},  {\em Phys. Rev. Lett.} {\bf
  117} (2016), no.~11 111601, [\href{http://xxx.lanl.gov/abs/1605.06098}{{\tt
  1605.06098}}].

\bibitem{Maldacena:2016upp}
J.~Maldacena, D.~Stanford, and Z.~Yang, {\it {Conformal symmetry and its
  breaking in two dimensional Nearly Anti-de-Sitter space}},  {\em PTEP} {\bf
  2016} (2016), no.~12 12C104, [\href{http://xxx.lanl.gov/abs/1606.01857}{{\tt
  1606.01857}}].

\bibitem{Engelsoy:2016xyb}
J.~Engelsoy, T.~G. Mertens, and H.~Verlinde, {\it {An investigation of
  AdS$_{2}$ backreaction and holography}},  {\em JHEP} {\bf 07} (2016) 139,
  [\href{http://xxx.lanl.gov/abs/1606.03438}{{\tt 1606.03438}}].

\bibitem{brezin1997spectral}
E.~Br{\'e}zin and S.~Hikami, {\it Spectral form factor in a random matrix
  theory},  {\em Physical Review E} {\bf 55} (1997), no.~4 4067,
  [\href{http://xxx.lanl.gov/abs/cond-mat/9608116}{{\tt cond-mat/9608116}}].

\bibitem{prange1997spectral}
R.~Prange, {\it The spectral form factor is not self-averaging},  {\em Physical
  Review Letters} {\bf 78} (1997), no.~12 2280,
  [\href{http://xxx.lanl.gov/abs/chao-dyn/9606010}{{\tt chao-dyn/9606010}}].

\bibitem{cotler2017black}
J.~S. Cotler, G.~Gur-Ari, M.~Hanada, J.~Polchinski, P.~Saad, S.~H. Shenker,
  D.~Stanford, A.~Streicher, and M.~Tezuka, {\it Black holes and random
  matrices},  {\em Journal of High Energy Physics} {\bf 2017} (2017), no.~5
  118, [\href{http://xxx.lanl.gov/abs/1611.04650}{{\tt 1611.04650}}].

\bibitem{maloneySFF}
A.~Maloney, {\it {To appear}}.

\bibitem{eynard2007invariants}
B.~Eynard and N.~Orantin, {\it Invariants of algebraic curves and topological
  expansion},  \href{http://xxx.lanl.gov/abs/math-ph/0702045}{{\tt
  math-ph/0702045}}.

\bibitem{murugan2017more}
J.~Murugan, D.~Stanford, and E.~Witten, {\it More on supersymmetric and 2d
  analogs of the syk model},  {\em Journal of High Energy Physics} {\bf 2017}
  (2017), no.~8 146, [\href{http://xxx.lanl.gov/abs/1706.05362}{{\tt
  1706.05362}}].

\bibitem{lin2020bootstraps}
H.~W. Lin, {\it {Bootstraps to strings: solving random matrix models with
  positivity}},  \href{http://xxx.lanl.gov/abs/2002.08387}{{\tt 2002.08387}}.

\bibitem{komargodski2017random}
Z.~Komargodski and D.~Simmons-Duffin, {\it {The random-bond Ising model in 2.01
  and 3 dimensions}},  {\em Journal of Physics A: Mathematical and Theoretical}
  {\bf 50} (2017), no.~15 154001,
  [\href{http://xxx.lanl.gov/abs/1603.04444}{{\tt 1603.04444}}].

\bibitem{eynard2015random}
B.~Eynard, T.~Kimura, and S.~Ribault, {\it {Random matrices}},
  \href{http://xxx.lanl.gov/abs/1510.04430}{{\tt 1510.04430}}.

\bibitem{stanford2019jt}
D.~Stanford and E.~Witten, {\it {JT gravity and the ensembles of random matrix
  theory}},  \href{http://xxx.lanl.gov/abs/1907.03363}{{\tt 1907.03363}}.

\end{thebibliography}\endgroup

\end{document}